\documentstyle[prd,preprint,aps,preprint]{revtex}
\begin{document}
\draft
\preprint{
\begin{tabular}{r}
JHU-TIPAC 95016
\end{tabular}
}
\title{
A Proposed Scale-Dependent Cosmology \\
for the Inhomogeneous Universe
}
\author{C. W. Kim\footnote{E-mail cwkim@rowland.pha.jhu.edu}
and J. Song\footnote{E-mail jhsong@rowland.pha.jhu.edu}}
\address{Department of Physics and Astronomy\\
The Johns Hopkins University\\
Baltimore, MD 21218, U.S.A.
}
\maketitle

\begin{abstract}
\setlength{\baselineskip}{.5cm}
We propose a scale-dependent cosmology in which the
Robertson--Walker metric and the Einstein equation
are modified in such a way that $\Omega_0$, $H_0$
and the age of the Universe all become
scale-dependent.
Its implications on the observational cosmology and
possible modifications of the standard Friedmann
cosmology are discussed.
For example, the age of the Universe in this model
is longer than that of the standard model.
\end{abstract}


\section{Introduction}
The standard Friedmann cosmology based on the Cosmological
Principle (or equivalently the Robertson--Walker metric)
and the Einstein equation have been very successful
in describing the Hot Big Bang Model of the Universe.
The present Universe, however, is not as simple
as the Universe once was in the early epoch.
The matter distribution that we observe now is
no longer homogeneous.
(Recall homogeneity of the Universe is one of the two
ingredients of the Cosmological Principle,
the other being isotropy of the Universe.)
That is, the homogeneous and isotropic
Universe, as described by the Robertson--Walker
metric, has evolved in time into the present inhomogeneous
Universe with galaxies, clusters, superclusters and other large
scale structures.
This means that in spite of its success, the standard model
is not suitable, at least in its original form,
for describing the present Universe with various structures, except
perhaps the global behavior of the visible Universe as a whole.
This has motivated us to modify the standard
model in order to describe the present inhomogeneous Universe.

First, since both the observed microwave background
radiation and the matter density distribution
appear almost isotropic to us at present epoch,
the isotropy of the Cosmological Principle appears to be valid
even at the present epoch.
Hence, in order to successfully describe the present Universe,
the homogeneity assumption may be modified or relaxed.
Furthermore, recent measurements of the Hubble constant,
$H_0$, using NGC 4571\cite{Pierce} (distance=14.9$\pm$1.2 Mpc)
and M100\cite{Freedman}(distance=17.1$\pm$1.8 Mpc)
in the Virgo cluster and the type Ia supernovae\cite{Sandage},
and their implied
ages of the Universe have become a subject of heated controversy.
In addition to the disagreement among the measurements,
the major problem is that the age of the Universe predicted
by the use of the larger value of $H_0$ from NGC 4571
($H_0=87 \pm 7$ Km/secMpc) or from M100 ($H_0 \simeq$80
Km/secMpc) and the standard Friedmann
cosmology with flat space becomes about half the
measured ages of 14 to 18 Gyr for the oldest stars and
globular clusters.
Clearly, both the measurements and
the standard cosmology cannot be right.

Another observational fact is that the observed matter density,
when viewed as a function of the cosmic scale
is definitely not a constant, at least up to
several hundred Mpc, not in accord with
the standard cosmology prediction.
Instead it is an increasing function of scale.\cite{Schramm}

On the theoretical side, it has recently been suggested that
the gravitational constant, $G$, may be asymptotically free in the
effective Asymptotically-Free-Higher-Derivative (AFHD) quantum gravity
\cite{Julve}.
(Some implications on the content of dark matter in the Milky Way
and the Universe  as well as on the evolution of the Universe
have been discussed in the past by various authors
\cite{Bottino}
.)
Although the running of $G$ is due to the quantum effects
at the Planck mass scale, it was suggested in Ref.[7]
that the running of $G$ may be prominent
in very large distance scales beyond 1 Kpc.
This puzzling aspect is not satisfactorily
understood, to say the least, at present.
For this presumed asymptotically-free behavior, see,
for example, figure 2 in \cite{Goldman} and
figure 1 of \cite{Bertolami}.
On the other hand,
a new scale-dependent cosmology in which
the Hubble constant, matter density and the present age of
the Universe all became scale-dependent was
recently proposed by relaxing the
Cosmological Principle and generalizing the Einstein equation
to accommodate the scale-dependence\cite{Kim}.

In this paper, we present a detailed study of
the scale-dependent cosmology proposed in \cite{Kim}.
The plan of our paper is as follows.
In section II, we present in detail the motivation
for relaxing the Cosmological Principle
and give a rationale for generalizing the Einstein equation
so that the scale-dependence can be accommodated.
Their consequences and comparison with the standard Friedmann cosmology
are given in section III.
Section IV is devoted to subtle features of the
new metric such as necessity of keeping the pressure
and the momentum density non-vanishing throughout even in the
matter-dominated era.
The behavior of the density, pressure and momentum density
in this cosmology as functions of time $t$ and coordinate parameter $r$,
and several possible cosmological scenarios are discussed in section V.
Section VI is devoted to some cosmological consequences of this model,
including how expansion rate, mass density and
the age of the Universe depend on the scale parameter $r$.
In this section, we also discuss the meaning of the
scale-dependence, in particular, what quantities are observable
or physical, and the nature of a constraint on the two parameters, time $t$,
and coordinate parameter $r$ when actual observations are made.
A summary and discussions are given in section VII.

\section{Modification of the Cosmological Principle}

The Cosmological Principle states that the Universe is isotropic
and homogeneous to any observer in the Universe at any given time.
If the Universe is isotropic to every observer, it is also
homogeneous\cite{Weinberg}.
For two observers who are separated by a comoving scale
vector $\vec{r}_0$, isotropy means
\begin{equation}
\rho(r) = \rho(\vec{r}) =\rho(\vec{r} + \vec{r}_0)~~.
\end{equation}
Since Eq.(1) holds for an arbitrary $\vec{r}_0$,
$\rho(r)$ has to be a constant.
That is, the Universe is homogeneous.
It is to be emphasized that the strict sense of the
Cosmological Principle can only be satisfied by a Universe
in which the energy-momentum tensor takes the form
of a perfect fluid, given by
\begin{equation}
T^{\mu\nu}
= (\rho + p)u^{\mu}u^{\nu} - p g^{\mu\nu} ~~,
\end{equation}
where $u^{\mu}$ is a {\it velocity four-vector \/}
with the components
\begin{equation}
u^0	=	1 ~~,~~
u^i	= 	0	~~~( ~i=1,2,3~ )~,
\end{equation}
and $\rho$ and $p$ are the energy density
and pressure, respectively\cite{Weinberg}.
In a Universe consisting of a perfect fluid, therefore,
the momentum density, $T_{0i}$, must vanish.

The Universe, however, has not been, strictly
speaking, homogeneous ever since the
density fluctuations were created during the inflation
in the very early epoch.
At the present epoch, galaxies are not
uniformly distributed over the Universe.
Rather, galaxies tend to cluster and many
clusters are again gravitated to form superclusters.
Moreover, based on the observation \cite{Schramm},
the mass density appears to increase with the cosmic scale.
We, therefore, wish to find a proper metric
which approximately describes this complex Universe and accommodates
the observed increase of the energy density
as a function of scale.

To this end, we make the first ansatz
that the Universe is described by the metric
\begin{equation}
d\tau^{2} = dt^2 - R^2(t,r)(dr^2 + r^{2}d\Omega^2)~~,
\end{equation}
where $R(r,t)$ is the generalized scale factor
which depends on $r$ as well as on $t$.
This metric manifestly violates homogeneity
in the Cosmological Principle
so that each observer sees different Universe.
As in the case of the Schwartzschild metric,
there is a special observer to whom the Universe
is perfectly isotropic.
We may or may not be this special observer.
However, since the observed cosmic microwave
background radiation and mass density (proportional
to $\Omega_0$) look almost isotropic to us,
we cannot be too far from this observer.
In this paper, for mathematical simplicity, we take our position
 as the origin of the coordinate system.
A similar view of our location in the Universe
(albeit the return of the pre-Copernican notion)
was proposed by Linde et. al. in a paper ``Do we live in the
center of the world ?''\cite{Linde}.

Now that the metric is specified, the dynamics of the Universe
is governed by the Einstein equation.
Since the metric in Eq.(4) introduces the $r$-dependence
in the Einstein tensors which appear on the left-hand side
of the Einstein equation,
the right-hand side proportional to $GT_{\mu\nu}$ has to be modified
so as to accommodate the $r$-dependence.
In any gravitational experiments,
$G$ cannot be measured separately from the
energy-momentum tensor.
Therefore we propose that the Einstein equation be generalized as
\begin{equation}
R^{\mu \nu}-\frac{1}{2}g^{\mu \nu}R=-8\pi [G T^{\mu \nu}](r,t)~,
\end{equation}
implying
\begin{equation}
[G T^{\mu \nu}]_{; \nu} =0,
\end{equation}
where the semicolon denotes a covariant derivative.
This is our second ansatz.
In Eq.(5), we have explicitly written down the $r$ and $t$ dependence
of $GT_{\mu\nu}$.
Moreover, $T_{\mu\nu}$ cannot be the energy-momentum
tensor of a perfect fluid.
Of course, we should recover the Einstein equation in the
limit of $r\rightarrow 0$ because
$G$ becomes the Newton's gravitational constant,
$G_N$, when $r=0$ and at present epoch $t =t_0$.

\section{Consequences of the generalized Robertson--Walker
 metric and Einstein equation}
Having generalized both the Robertson--Walker  metric and the Einstein
equation, we now proceed to discuss their consequences.
When the non-vanishing elements of Ricci tensor calculated
from Eq.(4) are substituted into the
generalized Einstein equation in Eq.(5),
we obtain the following non-vanishing components :
\begin{eqnarray}
3 \frac{\dot{R}^2(t,r)}{R^2(t,r)}-2\frac{R''(t,r)}{R^3(t,r)}
+\frac{R'^2(t,r)}{R^4(t,r)}-4\frac{R'(t,r)}{rR^3(t,r)}
&=&8 \pi [G\rho] \\
2 \frac{\ddot{R}(t,r)}{R(t,r)}+\frac{\dot{R}^2(t,r)}{R^2(t,r)}
-\frac{R'^2(t,r)}{R^4(t,r)}-2\frac{R'(t,r)}{rR^3(t,r)}
&=& -8 \pi [Gp_r] \\
2 \frac{\ddot{R}(t,r)}{R(t,r)}+\frac{\dot{R}^2(t,r)}{R^2(t,r)}
-\frac{R''(t,r)}{R^3(t,r)}
+\frac{R'^2(t,r)}{R^4(t,r)}-\frac{R'(t,r)}{rR^3(t,r)}
&=&- 8 \pi [Gp_{\theta}] \\
2 \frac{\ddot{R}(t,r)}{R(t,r)}+\frac{\dot{R}^2(t,r)}{R^2(t,r)}
-\frac{R''(t,r)}{R^3(t,r)}
+\frac{R'^2(t,r)}{R^4(t,r)}-\frac{R'(t,r)}{rR^3(t,r)}
&=&- 8 \pi [Gp_{\varphi}]
\end{eqnarray}
where dots and primes denote, respectively, derivatives with
respect to $t$ and $r$.
Another non-vanishing Ricci tensor $R_{01}$ yields
\begin{equation}
R_{01} = 2 \left( \frac{\dot{R}'(t,r)}{R(t,r)}
	-\frac{\dot{R}(t,r)R'(t,r)}{R^2(t,r)} \right)
	= - 8\pi [GT_{01}]~~.
\end{equation}
We keep $T_{01}$ to be finite as will be explained below.
In addition, even in the matter-dominated
era, a finite pressure must be kept throughout.
The detailed discussion on the role of non-zero pressure and $T_{01}$
is given in the section IV.
We emphasize here that when the $r$ dependence is simply $dropped$
(i.e., $R' = R'' =0$ ) and  $p_{r}=p_{\theta}=p_{\varphi}=p$
in Eqs.(7)--(11),
the Friedmann equations for $k$=0 are recovered.

Before we proceed,
we make a brief comment on the previous works
on the scale-dependent cosmology.
History of scale-dependent cosmology dates back
to as early as 1930's.
In order to study the gravitational evolution of large
structures such as clusters, the scale-dependent cosmology
was discussed independently by Lemaitre,
Dingle and Tolman \cite{Lemaitre}.
They considered an inhomogeneous but isotropic Universe as
in our case.
However, our model differs from theirs in a very significant way.
They all assumed $zero$ pressure and $zero$ radial
momentum density with a non-vanishing cosmological constant.
As we will discuss in the following section,
the vanishing pressure or $T_{01}$
in our metric given in Eq.(4) leads to the factorization
of the scale factor, $R(r,t)$, as $a(t)S(r)$,
in which case the Robertson--Walker metric is recovered.
In order to avoid this factorization,
Lemaitre, Dingle and Tolman introduced two different
scale factors for the radial and angular parts
in the metric, i.e.,
$d\tau^2=dt^2-[R_1^2(r,t)dr^2+R_2^2(r,t)d\Omega^2]$.
In this case, however,
the expansion rates at different scales were
dictated by different cosmological models
(see \cite{Peebles} for an excellent review.)
without being counter-balanced by changing
pressure and radial momentum density.
Furthermore, in their model, again due to the absence
of $p$, there appears a tidal field.
In our case, however, different expansion rates at
different scales are counter-balanced  by changing
energy density and non-zero pressure.
More recent study is typified by the work of Dyer and Roeder
\cite{Dyer}
in which inhomogeneity was treated as fluctuation to the standard
cosmology.

Even though $p_r$ appears to be different
from that of $p_{\theta}(=p_{\phi})$ as can be seen from
Eqs. (8) and (9),
we shall set them equal to each other.
This is justified from the vanishing of spatial
off-diagonal components of the Einstein tensors which are proportional
to energy-momentum tensors, implying no shear force.
Unless $p_r=p_{\theta}$, we have non-vanishing $r-\theta$
shear force proportional to $T_{12}$,
which is zero, leading to a contradiction.
Under the assumption that
$p_{r}=p_{\theta}=p_{\varphi}=p$,
therefore,
we find a constraint on $R(t,r)$ from Eqs.(8) and (9), given by
\begin{equation}
\frac{R''(t,r)}{R^3(r,t)}-2\frac{R'(t,r)^2}{R^4(t,r)}
-\frac{R'(t,r)}{rR^3(r,t)} =0~~.
\end{equation}
Let us rewrite Eq.(12) as
\begin{equation}
\frac{R''(t,r)}{R'(t,r)}-2\frac{R'(t,r)}{R(t,r)}=\frac{1}{r}~~,
\end{equation}
which can now easily be integrated, yielding
\begin{equation}
R(t,r)=\frac{a(t)}{1-\left[ {1+b(t) \over 4} \right]r^2}~~,
\end{equation}
where $a(t)$ and $b(t)$ are integration constants of
the $r$ integration,
i.e.,  arbitrary functions of $t$ alone.
We have chosen the definition of $b(t)$ for later convenience.
Moreover, we emphasize here again that
at least in the matter-dominated Universe
where scale-dependent mass density is expected,
$b(t)$ cannot be a constant;
otherwise, $R(t,r)$ is factored out.
Suppose that $[1+b]/4$ is a constant, $\kappa$,
then our metric is reduced to
\begin{equation}
d\tau^2 = dt^2 - a^2(t)\left[
\frac{dr^2}{(1-\kappa r^2)^2}
+\frac{r^2}{(1-\kappa r^2)^2}d\Omega
\right] ~~.
\end{equation}
Setting $\bar{r} = r/(1-\kappa r^2) $ leads to
\begin{equation}
d\tau^2 = dt^2 - a^2(t)\left[
\frac{d\bar{r}^2}{1+4\kappa \bar{r}^2}
+\bar{r}^2d\Omega \right],
\end{equation}
which is the Robertson--Walker metric for
$\kappa$ equal to $0$ or $\pm 1/4$,
which cannot accommodate the scale-dependent mass density
in the matter-dominated era.
We, therefore, do not consider the case of $R(t,r)$ being factored out.

The form of $R(t,r)$, Eq.(14),
allows us to simplify the cosmological equations.
When Eq.(14) is substituted into Eq.(7),
Eq.(7) becomes
\begin{equation}
\left[  \frac{\dot{R}(t,r)}{R(t,r)} \right]^2
=	\frac{8\pi }{3}[G\rho](t,r) + \frac{1}
{a^2(t)} + \frac{ b(t)}{a^2(t)}~~,
\end{equation}
and using Eq.(12), Eqs.(7) and (8) yield
\begin{equation}
\frac{\ddot{R}(t,r)}{R(t,r)}=-\frac{4 \pi}{3} [
	G\rho + Gp ]~~.
\end{equation}
Note that setting $r=0$ in Eq.(17) would not reproduce
the Friedmann equation with $k=0$ because of the
$[1+b(t)]/a^2(t)$ term.
Recall that we are $not$ allowed to set $r$
equal to zero from the very beginning.
Our metric in Eq.(4) can now be written as, with Eq.(14),
\begin{equation}
d\tau^2 = dt^2 - \left[\frac{a(t)}
{1- \left[ {1+b(t) \over 4} \right]
 r^2}\right]^2( dr^2 + r^2d\Omega)~~.
\end{equation}
Note that the metric at $r=0$ is ill-defined.
(There does not even exist its inverse metric, $g^{\mu\nu}$.
In the case of the Robertson--Walker metric,
the open or closed Universe
cannot be said to become flat one simply
because they have the same behavior at $r=0$.)
The physical meaning of Eq.(17) is as follows.
As we have repeatedly mentioned, as $r$ increases,
i.e., as we examine larger and larger scales,
the present value of $G\rho$ should increase,
as suggested by the observed increase of the matter density.
As the proper cosmic distance grows at any fixed time,
the $ G\rho$ term also grows,
whereas the $[1+b(t)]/a^2(t)$ term remains fixed.
Thus the last two terms on the right-hand side in Eq.(17) become
relatively  small compared with the $G\rho$ term.
This is the way we recover a flat Friedmann Universe at large distance.

\section{Why we need $T^{01} \not= 0$ and $p\not= 0$.}
\subsection{Non-Vanishing $T_{01}$}
Although its numerical value at present epoch may not be large,
the radial momentum density,
$T_{01}$, plays a vital role in allowing matter to flow,
due to gravity, into larger structures
so as to make the present Universe look like as it is
and to make $\Omega_{0}(r)$ an increasing function of $r$.
(Strictly speaking, neither $T_{01}$ nor $T^{01}$
is the physical momentum density, which is obvious from its dimension.
This is because our metric is not orthonormal.
Even though $\sqrt{T^{01}T_{01}}$ is a physical one,
the following discussion does not depend on this distinction.)
In addition,  it also prevents $R(t,r)$ from being factored out as
a product of $a(t)$ and $S(r)$.

In order to demonstrate this, we start with Eq.(11) with the
non-zero Ricci tensor, $R_{01}$, which has not been utilized so far.
If we take  $T_{01}=0$ in Eq.(11),
which is the case for a perfect fluid (Eqs. (2) and (3)),
then Eq.(11) forces $R(t,r)$ to be indeed factored out,
which is of no interest to us as mentioned before.
In order to see this, we use the following identity
\begin{equation}
\frac{\dot{R}'}{R}= \frac{1}{R}\frac{\partial}{\partial r}
                    (R \frac{1}{R}\dot{R})
     =  \frac{R'}{R} \frac{\dot{R}}{R} + \frac{\partial}{\partial
r}\frac{\partial}{\partial t}\ln R
     ~~.
\end{equation}
Substituting Eq.(20) into Eq.(11) and setting $T_{01}=0$ in Eq.(11),
we find
\begin{equation}
\frac{\partial}{\partial r}\frac{\partial}{\partial t}\ln R =0~~,
\end{equation}
which implies that $R(r,t)$ must be in the form of $a(t)S(r)$.
Therefore, as we have repeatedly mentioned, in order to describe
the Universe as we see it  now by using our new metric,
it is necessary to have imperfect fluid
with $T_{01} \not= 0$, which allows matter to flow from the
originally almost
homogeneous distribution to the present inhomogeneous distribution.
This
very condition, $T_{01} \not =0$, also protects the
form of $R(t,r)$ as given in Eq.(15) from being factored out.

Another reason why $T_{01}$ should not be zero is as follows. The
conservation of the new energy-momentum tensor, $GT^{\mu \nu}$, is
expressed as
\begin{equation}
[GT_{\mu}^{\nu}]\,_{;\nu}
=[GT_{\mu}^{\mu \nu}]\,_{,\nu}
- \Gamma_{\mu\nu}^{\sigma}(GT_{\sigma}^{\nu})
+ \Gamma^{\nu}_{\nu\sigma}(GT_{\mu}^{\sigma}) =0 ~~.
\end{equation}
The $\mu=1$ component of Eq.(22) with vanishing $T_{01}$ becomes
\begin{equation}
0= (GT_1^{\nu})_{;\nu}
 =(GT_1^{\nu})_{,\nu} - \Gamma^\sigma_{1\nu}(GT_{\sigma}^{\nu})
     + \Gamma^{\nu}_{\nu\sigma}(GT_1^{\sigma})
\end{equation}
Substituting $T_1^1=T_2^2=T_3^3=-p(r,t)$,
Eq.(23) simply reduces to
\begin{equation}
(Gp)'=0 ~~,
\end{equation}
which implies that $Gp$ has no $r$ dependence.
Therefore, $T_{01}=0$ imposes the constraint on $Gp$ that $Gp$ be only
time-dependent. If this were true,
since $G$ is conjectured here to have an
$r$ dependence, this $r$
dependence is miraculously cancelled by that of $p$, which is
highly unlikely.
Hence, we must keep $T_{01}$ in the new energy-momentum
conservation equation in Eq.(22) in order to obtain physically
consistent results.
We have demonstrated that the non-vanishing of $T_{01}$ is
essential not only to
prevent $R(t,r)$ from being factored out
but also to keep the $r$ dependence
in $Gp$ which is a characteristic of an imperfect fluid.

\subsection{Non-Vanishing Pressure}
As was discussed in the Ref. \cite{Kim}, the scale dependence of
$(G\rho)$ is essential in the matter-dominated era
in order to have the inhomogeneous
matter distribution.
In the standard Friedmann Cosmology where the Universe is considered as
consisting of a perfect fluid, the pressure can be approximated to be
zero in the matter-dominated era.
This, however, is not the case in our new cosmology, at least
for mathematical consistency.
Let us examine the consequence of setting the pressure
equal to zero in our model.

First, recalling the general solution of $R(r,t)$ in Eq.(14),
we will show that $b(t)$ becomes a constant unless
$p_r=p_{\theta}=p_{\phi}\not=0$.
Supposing that all pressure terms are exactly zero,
we have, from Eq.(18),
\begin{equation}
\frac{\ddot{R}}{R} + \frac{4\pi}{3}(G\rho)=0~.
\end{equation}
Using the above equation,
we can eliminate $G\rho$ term in Eq.(17), yielding
\begin{equation}
2\frac{\ddot{R}}{R}
+ \left(\frac{\dot{R}}{R}\right)^2 = \frac{1}{a^2}+\frac{b(t)}{a^2}~~.
\end{equation}
We now calculate the left-hand side of
the above equation using Eq.(14) as
\begin{eqnarray}
2\frac{\ddot{R}}{R}+ (\frac{\dot{R}}{R})^2 &=&
\frac{1}{a^2(1-Br^2)^2}\{(2a\ddot{a}+\dot{a}^2)
                     \nonumber \\
          & & +r^2( 6a\dot{a}\dot{B}+2a^2\ddot{B}
                    -2\dot{a}^2B-4a\ddot{a}B)  \nonumber \\
          & & +r^4( 2a\ddot{a}B^2+\dot{a}^2B^2+5a^2\dot{B}^2
                    -2a^2B\ddot{B}-6a\dot{a}B\dot{B})~\}
                     \nonumber \\
          &\equiv& \frac{1}{a^2(1-Br^2)^2}\{C_0(t)+C_2(t)r^2
                    +C_4(t)r^4\}~~,
\end{eqnarray}
where $B(t)$ denotes $[1+b(t)]/4$ for simplicity.
Equating the above expression to the right-hand side of Eq.(26)
which has no $r$-dependence, we find the following three conditions
\begin{eqnarray}
2 a \ddot{a} + \dot{a}^2        &=&     4 B    \nonumber \\
3a\dot{a}\dot{B}+a^2\ddot{B} &=& 0        \nonumber \\
5a^{2}\dot{B}^{2}-2a^2B\ddot{B}-6a\dot{a}B\dot{B} &=& 0 ~~ .
\end{eqnarray}
The last two equations have a unique solution $\dot{B}=\dot{b}/4=0$,
implying that $R(t,r)$ is factored out as $a(t)S(r)$.
(The first equation is nothing but
a Friedmann equation, if we take $B(t)$ to be $1/4$.)
Therefore, no matter how small they are compared
with the mass density term even in the matter-dominated era,
it is essential to keep the pressure terms throughout,
for mathematical consistency.

\section{Behavior of $G\rho$, $Gp$ and $GT_{01}$}

We are now ready to discuss how the matter density,
pressure and momentum density behave in this new metric.
We first note that the constraint on $R(t,r)$ severely restricts
the behavior of $G\rho$, $Gp$ and $GT_{01}$
as functions of $t$ and $r$.
Substituting Eq.(14) into Eqs.(7), (8) and (11),
we obtain the following explicit expressions :
\begin{eqnarray}
\frac{8\pi}{3}G\rho &=&
	\left(\frac{\dot{a}}{a} \right)^2
	-\frac{1}{a^2}-\frac{ b}{a^2}
	+ 2\left(\frac{\dot{a}}{a}\right)\left( \frac{\dot{b}r^2/4}
	{1- \left[ {1+b \over 4} \right]r^2}\right)
		+\left(\frac{\dot{b}r^2/4}
	{1- \left[ {1+b \over 4} \right]r^2} \right)^2
			 \\
8\pi Gp&=&
	\frac{1}{a^2}-
	2\frac{\ddot{a}}{a} - \left( \frac{\dot{a}}{a}\right)^2
	+\frac{b}{a^2}
	- \left( 6 \frac{\dot{a}}{a}+
	2\frac{\ddot{b}}{\dot{b}} \right)
\left(	\frac{\dot{b}r^2/4}
	{1- \left[ {1+b \over 4} \right]r^2}\right)
	 - 5 \left(
	\frac{\dot{b}r^2/4}
	{1- \left[ {1+b \over 4} \right]r^2} \right)^2
			\\
8\pi GT^{01} &=& \frac{\dot{b}r}{a^2} ~~,
\end{eqnarray}
where $a(t)$ and $b(t)$ are completely arbitrary functions.
Although we can take $a(t)$ to be positive because it
becomes the scale factor at $r\simeq0$,
we defer the determination of the sign of $1+b(t)$.

Before we proceed to discuss physical constraint
on $a(t)$ and $b(t)$,
some comments on the apparent singularity in Eq.(29)
and (30) are in order.
When $[1+b(t)]>0$,
we encounter the singularity in $G\rho$ and $Gp$
at $r=2/\sqrt{1+b}$.
This singularity, however, is not a true singularity
but rather a spurious one, as in the case of $r=1$
for the Robertson--Walker metric with $k=1$
(Recall $d\tau^2=dt^2-R^2(t)[{dr^2 \over 1-kr^2}+
r^2d\Omega^2]$.)
Similarly, $r=2MG$ in the Schwartzschild metric is not a true
singularity although the factor $1/(1-2GM)$ in front of
$dr^2$ becomes infinity at this point.
(One can transform away this infinity
by a proper coordinate transformation.)
The position $r=2GM$ is in fact the horizon.
The true singularity in the above two examples is located
where the spatial scalar curvature diverges,
which is the case at $R(t)=0$ or $t=0$
in the Friedmann cosmology and at $r=0$
in the Schwartzschild metric.
In our metric, the spatial curvature $R^{(3)}$ is given by
\begin{eqnarray}
R^{(3)} &=&
 \frac{1}{R^2(t,r)} \left(
	\frac{R'^2(t,r)}{R^2(t,r)}-2\frac{R'(t,r)}{rR(t,r)} \right)
\\ \nonumber
&=& -\frac{1+b(t)}{a^2(t)}~~,
\end{eqnarray}
where we have used Eqs.(12) and (14).
Therefore, $R^{(3)}$ is scale-independent
and is always finite except when $a(t)=0$ at $t=0$.
That is, the singularity at $r=2/\sqrt{1+b}$ is
spurious and we interpret this position
as the horizon in our metric.
The sign of the spatial scalar curvature is, of course,
determined by the sign of $1+b(t)$.

Let us check if the energy-momentum conservation
equation, Eq.(22), that
consists of four independent components,
is satisfied by Eqs.(29)--(31).
If not, it implies further constraint on $a(t)$ and $b(t)$.
Under the condition that the pressure is a function of $r$ and $t$
and $T^1_1=T^2_2=T^3_3=-p(r,t)$,
the two components for $\mu=2$ and $\mu=3$
in Eq.(22) are automatically satisfied.
Equation (22) for $\mu=0$ is
\begin{equation}
\frac{\partial}{\partial t}[G\rho] + 3\frac{\dot{R}}{R}[G\rho]
+ \frac{\partial}{\partial r}[GT^1_0]
+ ( 3\frac{R'}{R}+\frac{2}{r} )[GT^1_0]
+ 3 \frac{\dot{R}}{R}[Gp] = 0 ~~,
\end{equation}
which is trivially satisfied when
$G\rho$, $Gp$ and $GT_{01}$ given in
Eqs.(29)--(31) are substituted, implying self-consistency.
For $\mu=1$, we have
\begin{equation}
-[Gp]' + \frac{\partial}{\partial t}[G T^0_1]
+ 3 \frac{\dot{R}}{R}[GT^0_1] = 0 ~~,
\end{equation}
which can be reproduced by taking $r$-derivative of Eq.(8).
Therefore, the energy-momentum conservation does $not$
impose any new constraint on $a(t)$ and $b(t)$.
This is in fact no surprise since Eqs.(29)--(31)
 were obtained with the use of the
Einstein equation which has the energy-momentum
conservation built into it.
The evolution of the Universe is,
therefore, completely determined by the
as-yet unknown two functions of $t$, $a(t)$ and $b(t)$.
Of course, if $G\rho(t,r)$, $Gp(t,r)$ and $GT_{01}(t,r)$
are known (or given), $a(t)$ and $b(t)$ are determined.
As of now, we lack this information
except for the following physically plausible constraint
based on the expected behavior of the Universe.

In the very early epoch, we expect a homogeneous
radiation-dominated Universe.
(For practical purposes, we can ignore the
primordial density fluctuations here.)
Therefore, before the radiation-matter cross-over time, $t_{co}$,
we must have
\begin{equation}
\dot{b}(t) \simeq 0 ~~~~ \mbox{for} ~~~~t<t_{ {co}}~~,
\end{equation}
which is required from the presumed
homogeneity ($r$-independence) as can
be seen in Eq.(29).
If the pressure is related to
the mass density by $3p\simeq \rho$
in the radiation-dominated era,
we have, from Eqs.(29) and (30),
\begin{equation}
\left( \frac{\dot{a}}{a} \right)^2 - \frac{1+b}{a^2} \simeq
- \frac{ \ddot{a}}{a} ~~~~ \mbox{for} ~~~~ t<t_{{co}}.
\end{equation}
Even though we do not have simple power-law (in $t$) expressions for
$a(t)$ and $b(t)$ to explain the entire history of the Universe
(which is also the case of the closed or
open Universe in the Friedmann model),
if we restrict our discussion to the very early epoch, we could
try $a(t)$ in the form of $t^m$.
Since Eq.(35) implies a constant $b(t)$ in the very early epoch,
if $m$ is less than one,
the first term on the left-hand side of  Eq.(36),
proportional to $1/t^2$,
is dominant over the second term in the early epoch.
Therefore, the trial function,
$a(t) \sim t^{\frac{1}{2}}$, as  in the
Friedmann Cosmology in the radiation-dominant era,
approximately satisfies Eq.(36).
As far as the early Universe is concerned, therefore,
the evolution is almost identical to that of the Friedmann Cosmology,
if we impose $b(t)$ to be a constant in the early epoch.
This is all we can say about $a(t)$ and $b(t)$
in the early Universe.
We note that even the sign of $1+b(t)$ remains
undetermined since the behavior of the early Universe
is insensitive to the curvature term.

As the Universe evolves into the matter-dominated era,
inhomogeneity starts to appear,
which is generated by $\dot{b} \neq 0$,
as can be seen in Eqs.(29) and (31).
Its physical origin can be attributed, at least in our picture,
to the non-zero radial momentum-density (see Eq.(31)).
Since we expect the mass density to increase with the scale
in the matter-dominated era, as is seen from the observation,
we require that
\begin{equation}
\dot{b}(t) >0 ~~\mbox{for}~~t \gg t_{ {co}}~~ .
\end{equation}
Recalling that $G\rho (t,r \simeq 0)=(\dot{a}/a)^2-(1+b)/a^2$,
this increasing $b(t)$
makes our local neighborhood relatively less
dense than in the case of the Friedmann cosmology with $k=0$,
which is expected.
Even though $b(t)$ was completely arbitrary,
the above  plausible physical assumptions allow us to
classify its qualitative behavior
into the following three cases,
based on the possible singularity in $R(t,r)$
depending on the sign of $b(t)$.
First, $1+b(t)$ starts from a $non-negative$ constant
and keeps increasing (case 1).
Secondly, $1+b(t)$ starts from a $negative$ constant,
has an increasing period and finally converges to zero (case 2).
Finally, even though $1+b(t)$ starts from a $negative$ constant,
it keeps increasing, eventually becoming positive (case 3).
Let us examine the cosmological consequences of each case.

\begin{enumerate}
\item
In this case, as soon as $b(t)$ starts growing,
the horizon or a shell with apparently infinite mass density
appears at large scale ( at $r=2/\sqrt{ 1+b(t)}$
as can be seen in Eq.(29)),
suggesting a picture in which our Universe is inside a
bubble or cavity with a wall
with a very high mass density (or the event horizon).
That is, in this case, $b(t)$ must remain zero in the early epoch and
after the cross-over and recombination times,
it starts to grow. At this moment the horizon appears.
Since $b(t)$ is an increasing function of $t$
as the Universe evolves
(see the denominator in Eq.(29)),
the horizon approaches us, located at or near the center of the
Universe ($r \simeq 0 $), with the velocity
proportional to $\sqrt{\dot{b}(t)}/[1+b(t)]^{3/2}$.
It is natural to expect that this incoming velocity
will decrease to zero eventually.
That is, $b(t)$ starts out at zero and remains so in the early Universe
and then it starts to grow and eventually flattens
to be a constant again.
This picture can be made consistent with the absence
of the measurements of such a shell or wall,
because it is always possible to push it outside   the
visible horizon by properly choosing numerical values of
$b(t)$ at present epoch.

\item
In this case, $1+b(t)$ is always negative so that
there exists no singularity in $R(t,r)$.
In addition, the mass density converges to a finite value
at any given time, as $r$ becomes large.
Therefore, we have following overall picture
of mass density in this case.
In the very early epoch, we have the homogeneous mass distribution.
After $t_{co}$, as $b(t)$ starts growing,
$G\rho$ decreases in time faster than in the standard Friedmann cosmology
with $k=0$ in our neighborhood.
At any given time, $G\rho$ is an increasing function of scale,
converging to a certain value.
Since $b(t)$ is an increasing function,
the limiting point, $r_c$,
where the asymptotic limit is effectively reached,
approaches us as time evolves (see the last term
in Eq.(31)).
And the asymptotic value of $G\rho$,
at infinite $r$, decreases in time
as can be seen in Eq.(29).

In spite of its reasonable appearance,
the case 2 has a serious defect.
The Universe was supposed to be flat ($k=0$)
and homogeneous in the very early epoch,
and as $b(t)$ starts growing,
the non-zero momentum density causes the mass to flow
 out from the center of the Universe.
Therefore, at present epoch,
it is expected that the Universe is locally $open$ and
a flat one is recovered at large $r$.
In the case 2, however, the spatial curvature factor,
$-[1+b(t)]/a^2(t)$ which is $r$-independent, is positive,
implying the  $close$ Universe.
(Of course, since $[1+b(t)]$ goes to zero as $t$ goes to infinity
while $a(t)$ keeps increasing,
the spatial curvature of the Universe eventually becomes zero
in the far future.
{}From the observed increase of mass density with scale,
if the case 2 is valid, $b(t)$ appears to be in the increasing phase,
implying that our present
Universe has a positive curvature.)
Regardless of this shortcoming, the case 2 is interesting
in  view of its overall picture of the Universe.

\item
This is a combination of case 1 and case 2 discussed above.
At the instant $[1+b(t)]$ changes its sign,
the horizon appears at $r=2/\sqrt{1+b}$.
However, there is very little information, for example,
on when the horizon appears.
And in the case 3, we have a Universe with varying curvature,
while the signs of curvature in both case 1 and 2
do not change in time.
That is, as the Universe evolves, the originally closed Universe becomes
open and eventually evolves into a flat one.
\end{enumerate}

A final but very important comment in this section is
on the behavior of the pressure.
Regardless of the sign of $[1+b(t)]$ and $\dot{b}(t)$,
the pressure $p$ becomes negative beyond a certain value of $r$,
although it may be positive in our neighborhood.
Physical origin of this behavior can be traced back to the fact
that the new metric  allows a vacuum energy density
or the time-dependent cosmological constant.
In a special case of $b(t) \sim a^2(t)$,
the last term in Eq.(17) becomes a constant,
mimicking the cosmological constant.
Recently, cosmological models with $t$-dependent
$\Lambda$-term have been discussed by several authors
\cite{Ozer} in an attempt to explain
the dramatic discrepancy between the values of $\Lambda$
from the current observation ($\leq 10^{-55} {\mbox cm}^{-2}$)
and from the particle physics expectation which is higher by
a factor of  about $10^{120}$.
In these models, $\Lambda$ decreases
as the Universe evolves, since the vacuum energy
was necessary to create inflation and then particles.
In our scale-dependent cosmology, however,
a term which plays the role of the time-dependent
$\Lambda$-term is naturally generated by the
metric, whereas in Ref.[15] it was put by hand in the Einstein equation.

\section{Cosmological Consequences}
In this section, we present some details
on the scale-dependence of cosmological quantities.
Before we proceed, however, we must caution the reader
that the scale-dependent cosmological quantities
such as $H(t,r)$, ${\cal H}(t,r)$, $\Omega(t,r)$
to be discussed below are $not$ observable.
Since they represent values at time $t$
and at the coordinate parameter $r$,
except the values at $t=t_0$ and $r=0$
(i.e. present local value), observation  of them cannot be made
by the conventional
light propagation which has a finite speed.
Observational (or physical) interpretation of these
quantities will be discussed at the end of this section.

First, the Universe is expanding with the characteristic
scale factor, $R(t,r)$.
We begin by noting that we can define two expansion rates.
The first is the $proper$ expansion rate  defined by
\begin{equation}
H(t,r)=\frac{\dot{D}(t,r)}{D(t,r)} ,
\end{equation}
where the proper distance $D(t,r)$ is
\begin{equation}
D(t,r) \equiv \int _0 ^r  R(t,r') dr'.
\end{equation}
The second is
\begin{equation}
 {\cal H}(t,r) \equiv \frac{\dot {R}(t,r)}{R(t,r)},
\end{equation}
which we shall call the $scale$ expansion rate
for distinction and this is the one
that appears in our cosmological equation (Eq.(17)).
{}From  the explicit form of  $R(t,r)$ given in Eq.(14),
we can obtain the exact expressions for
$H(t,r)$ and ${\cal H}(t,r)$ in terms of $a(t)$ and $b(t)$.
First, the proper expansion rate,
$H(r,t)$, is given by
\begin{eqnarray}
H(t,r)	&=&\left(	\frac{\dot{a}(t)}{a(t)}
	-\frac{\dot{b}(t)}{2[1+b(t)]} \right)
\\ \nonumber
& &	+ \frac{\dot{b}(t)r/4}{2\sqrt{1+b(t)}
	[1-\left( \frac{1+b(t)}{4} \right)r^2]}
	\frac{1}{\mbox{tanh}^{-1}(\sqrt{1+b(t)}r/2)}~,  ~~
	\mbox {for} ~[1+b(t)]>0~,
\end{eqnarray}
\begin{eqnarray}
H(t,r)	&=&\left(	\frac{\dot{a}(t)}{a(t)}
	-\frac{\dot{b}(t)}{2[1+b(t)]} \right)
	\\	\nonumber
	& & - \frac{\dot{b}(t)r/4}{2\sqrt{-([+b(t)]}
	[1-\left( \frac{1+b(t)}{4} \right)r^2]}
	\frac{1}{\mbox{tan}^{-1}(\sqrt{-[1+b(t))]r/2)}}~
	,~~\mbox {for} ~[1+b(t)]<0~.
\end{eqnarray}
where for $[1+b(t)]>0$ we consider $r$ only in the range of
$0<r\sqrt{1+b(t)}/2 < 1$, which is inside the wall.
The scale expansion rate, ${\cal H}(t,r)$,
is simply given by
\begin{equation}
{\cal H}(t,r)=\frac{\dot{a}(t)}{a(t)}
		+ \frac{\dot{b}(t)r^2/4}
	{[1-\left( \frac{1+b(t)}{4} \right)r^2]}~~.
\end{equation}
Since the third terms on the right-hand sides of Eqs.(41) and (42)
both become $\dot{b}(t)/2[1+b(t)]$ as $r \rightarrow 0$,
cancelling the second terms,
the difference between $H(t,r)$ and
${\cal H}(t,r)$ disappears as $r\rightarrow 0$, i.e.,
\begin{equation}
H(t, r \simeq 0)={\cal H}(t, r \simeq 0)=\frac{ \dot{a}(t)}{a(t)}~~.
\end{equation}

The next question is whether the Universe has enough
mass to be closed or not,
which is characterized by the mass density parameter $\Omega$.
In the Friedmann Cosmology, the present value
$\Omega$ is defined by the ratio of
$\rho_0$ and the critical density, $\rho_{c,s}$,
\begin{equation}
\Omega_0 = \frac{\rho_0}{\rho_{c,s}}
=\frac{8 \pi G_N\rho_0/3}{H_0^2}~~,
\end{equation}
which is universal (independent of the locations) at any given time.
Here and hereafter, the subscript zero denotes a value
evaluated at the present epoch.
In our cosmology,  since both expansion rate and
mass density are functions of $r$,
the density parameter, $\Omega(t,r)$, is defined by
\begin{eqnarray}
\Omega(t,r) &\equiv&
\frac{[G\rho](t,r)}{[G\rho]_c(t,r)}
\\ \nonumber
&=&1-\frac{1+ b(t)}{a^2(t)}\frac{1}{{\cal H}^2(t,r)}~~,
\end{eqnarray}
where we defined $[G\rho]_c(t,r)$ as ${\cal H}^2(t,r)$.
Hence the observed $local$ values of
$\Omega(t,r\simeq0) \equiv \overline{\Omega}(t)$
and $H(t,r\simeq 0) \equiv \overline{H}(t)$
specify the value of $[1+b(t)]/a^2(t)$,
which is $[1-\overline{\Omega}(t)]\overline{H}^2(t)$.
Equation (46) can be rewritten as

\begin{equation}
\Omega(t,r)	=
1-[1-\overline{\Omega}(t)]
\left[1+ \left( \frac{a}{\dot{a}} \right)
	\frac{ \dot{b} r^2/4}
	{[1-\left( \frac{1+b}{4} \right)r^2]}   \right]^{-1} ~~.
\end{equation}
For $1+b(t)>0$, $\Omega(t,r)$ approaches unity
as $r$ approaches $2/\sqrt{1+b(t)}$.
At any given time, therefore, by setting $\overline{\Omega}(t)$
to be less than one, we have the local open Universe.
As we approach the horizon, the mass density
increases to make the Universe flat.
It is to be noted that $\Omega(t_0,r)$ is indeed the present value
of $\Omega$ as a function of $r$ but there is no way to measure
this now!  That is, $H(t_0,r)$, ${\cal H}(t_0,r)$ and $\Omega(t_0,r)$
are all $unobservable$ quantities in which $t_0$ and $r$ are independent.

We now discuss observational consequences of this scale-dependent
cosmology.
It is important to emphasize again that although
$t$ is the true cosmic time, the coordinate parameter $r$
is not a measurable quantity.
Physical quantities such as redshift $z$,
luminosity distance $d_{L}$,
mass density $\rho$ and so on are always measured via the light
which travels with a finite speed.
In order to make connection with observational data,
we must begin with and solve the light-propagation
equation or the null geodesic in this metric.
It is given by Eq.(19) with $d\tau=0$ as
\begin{equation}
dt	=	-{a(t) \over 1-\left[\frac{1+b(t)}{4}\right]r^2} dr
\end{equation}
where we consider $radial$ propagation of signal
(i.e. $\theta=\phi=const.$)
and the minus sign is taken so that $r$ decreases as $t$ increases.
Equation(48) is a first order differential equation whose
boundary condition is $r(t=t_0)=0$.
In the following, we mean by $r(t_0,t)$ the
comoving distance traveled by the light which
departed from the source at time $t$ and reaches us at time $t_0$,
implying the boundary condition $r(t_0,t=t_0)=0$.
With given $a(t)$ and $b(t)$, therefore,
the solution of the above differential equation
yields $r$ as a function of $t$ for the light ray
which we observe.
In the previous expressions for the mass density,
expansion rate and so on, $r$ was an independent comoving
coordinate.
However, whenever one determines  a physical quantity by $measurement$,
$r$ is no longer an independent coordinate.
That is, physical quantities $measured$ via the
$light$ propagation depend only on $t$,
even though they have explicit $r$-dependence in this cosmology.

Then, what is the meaning of the observed increase
of $\Omega$ as we look further out?
Whatever signal we are getting right now contains
information about the past in time.
In the standard cosmology, we $deduce$
physical quantities at the present epoch
by using the evolution law in the standard cosmology.
Measured are the red shift and $G\rho(t)$, not $G\rho(t_0)$.
For example, in the matter dominated era of the Friedmann cosmology
where the pressure is negligible,
$G\rho$ is proportional to $1/S^3(t)$.
Here $S(t)$ is the scale factor in the Friedmann cosmology
and $S(t_0)/S(t)$ is just  $1+z$.
In the framework of the Friedmann cosmology,
the $calculated$ $G\rho(t_0)$ is
the $observed$ $G\rho(t)$ times $1/(1+z)^3$.
Hence, the observed $\Omega^{obs}_0$ is
\begin{equation}
\Omega^{obs} \equiv {G\rho^{obs}(t_0) \over G\rho_{c,s} }
\equiv \frac{ G\rho(t,r(t_0, t))}{G\rho_{c,s}(1+z)^3}
\end{equation}
where $\rho_{c,s}$ is the critical density in the standard
cosmology.

Therefore, phenomenology of this scale-dependent cosmology depends
crucially on the solution of the light-ray equation in Eq.(48).
Since in Eq.(48)  the variables $t$ and $r$
 cannot be separated,
it is, in general, non-trivial to get an analytical solution
for $r(t)$.
Even for numerical evaluations,
we need functional forms of $a(t)$ and $b(t)$.
First, we assume that $b(t)$ can be approximated by
\begin{equation}
b(t)	=	\beta^2 \left(\frac{t}{t_0}\right)^n~~.
\end{equation}
(We consider only case 1 because of the locally open Universe.)
Secondly, Eqs.(29)--(31) evaluated at $r \simeq 0$ imply that
$a(t)$ is more or less the scale factor of
the locally open Friedmann Universe.
Therefore, it is reasonable to assume the followings properties of $a(t)$
\begin{equation}
2 \frac{ \ddot{a}(t) }{ a(t) }
+ \left[ \frac{ \dot{a}(t) }{a(t)} \right]^2-\frac{1}{a^2(t)} \simeq 0 ~~,
\end{equation}
which is motivated by the vanishing pressure
of the Friedmann cosmology in the matter-dominated era.
Solving Eq.(48) numerically with Eqs.(50) and (51), we can recast $r(t)$,
for example, in the form of
\begin{equation}
r(t_0,t)=\delta[ t_0^{\gamma}-t^{\gamma}]~~,
\end{equation}
where $\delta$ and $\gamma$ depend on
the numerical values of $\overline{\Omega}_0$, $\beta$, $n$ and so on.
The form of $r(t_0,t)$ in Eq.(52) was motivated by the fact that setting
$\gamma ={1 \over 3}$ in Eq.(52) yields the relationship between $t$
and $r$ in the Friedmann cosmology with $k=0$.
Recalling the meaning of $r(t_0,t)$ as discussed below Eq.(48),
$r(t_0,t)$ is equal to
$r(t_0+\Delta t_0,t+\Delta t)$ for two successive wave crests emitted
by a light source at $r$.
{}From this equality
we obtain a relationship between $z$ and $t$ as
\begin{equation}
1+z \equiv {\Delta t_0 \over \Delta t}
= \left({ t \over t_0}\right)^{\gamma -1}~~.
\end{equation}
Substituting the above  into Eq.(52),
we find
\begin{equation}
r(t_0,z)	=	{\delta \over \alpha}
[1-(1+z)^{\gamma \over \gamma-1}]~~.
\end{equation}
With Eqs.(53) and (54), $\Omega_{obs}$ can be expressed in terms
of the $measurable$ red-shift.
Finally, in order to extract physically meaningful expansion rates,
we need to know the relationship between the luminosity distance,
$d_L$, and $z$ and $r$. With this relation
and Eq.(54), we can obtain a modified Hubble's law which can be
confronted with the observation.
Unfortunately, however, because of the complexity of our metric,
we have so far  been unable to derive such a relationship but
in principle such a relation can be obtained. The actual derivation
is beyond the scope of this paper and it will be given elsewhere.
The only physically meaningful result of this scale-dependent
cosmology that can be derived at the moment is the age of the Universe. We will
discuss it in the following section.
\section{The Age of the Universe}

In the standard Friedmann cosmology,
often called the two-parameter model,
the present age of the Universe, $t_0$,
is uniquely determined by, for example,
$H_0$ and $\Omega_0$.
The well-known expression for $t_0$, valid for $\Omega_0 \sim 1$, is
\begin{equation}
t_0 \equiv \int_0^{R(t_0)} \frac{dR(t)}{\dot{R}(t)}
\simeq {1 \over H_0}\frac{1}{1+ \frac{1}{2} \Omega_0}~~.
\end{equation}
Since $H_0$ and $\Omega_0$ are scale-independent,
the age $t_0$ is location-independent.

In the scale-dependent cosmology under consideration here,
we have an additional parameter $r$ in all cosmological equations.
Hence, if any cosmological equation evaluated at present epoch
is solved for $t_0$,
it naturally becomes a function of $r$ as well as $H_0(r)$ and
$\Omega_0(r)$ (or the local values $\overline{H}_0$ and
$\overline{\Omega}_0$ and $r$), i.e., it appears that
the present age of the Universe depends on where we measure it.
The value $t_0(r\simeq 0) \equiv \overline{t}_0$
is the $local$ age of the Universe
as well as the true cosmic age, appropriate for
comparison with the age of the oldest stars and globular
clusters in our Milky Way.

When $a(t)$ and $b(t)$ in Eq.(17) or Eq.(29)
are known functions of $t$, then for example,
first setting $t=t_0$ in Eq.(29) and solving it for $t_0$
with use of the definitions of $\overline{H}_0$ and
$\overline{\Omega}_0$,
we can, in principle, obtain
\begin{equation}
t_0(r)=f(\overline{H}_0 ,  \overline{\Omega}_0, r)~~,
\end{equation}
where $f(\overline{H}_0$ ,  $\overline{\Omega}_0, r)$
is known when $a(t)$ and $b(t)$ are given.
In the Friedmann cosmology with $k=0$, this procedure indeed
yields the well-known result, $t_0=2/3H_0$.
The true age of the Universe is given by
$t_0(r=0)=f(\overline{H}_0, \overline{\Omega}_0, r=0)$.
Since the light propagation constraint on $r$ is yet to be
imposed, the age  $t_0(r)$ represents the present  age of the Universe
at $r$, which can be $measured$ only in the future.
Unfortunately, however, $a(t)$ and $b(t)$ are
presently unknown so that $f$ remains unknown.
In order to further elucidate the scale-dependent
(or location-dependent) nature of the age of the Universe,
we make the following crude approximations.
As in the case of the Friedmann cosmology,
we assume that the time duration of the matter-dominated era
up to now approximately represents the present age
so that in the energy-momentum conservation equation, Eq.(33),
the terms involving $G\rho$ are dominant over those
with $Gp$ and $GT_{01}$ for small $r$ as far as time-evolution is concerned.
Even though $Gp$ has a finite value at $r \simeq 0$,
the assumption in Eq.(51) and a proper choice of $b(t)$
can make it smaller than $G\rho$.
(Admittedly,  this is a poor approximation in our model,
but it serves as a working approximation in order to obtain a crude estimate.)
We then have, from Eq.(33),
\begin{equation}
G\rho \sim \frac{ 1 + \delta(r)}{R^3(t,r)}~~.
\end{equation}
The additional $r$-dependence denoted by $\delta(r)$
represents the correction to the standard result.
Defining, for a $fixed$ value of $r$,
Eq.(17) can now be written  with the definition,
\begin{equation}
x \equiv \frac{ R(t,r)}{R(t_0,r)}~~,
\end{equation}
as
\begin{equation}
\dot{x} = \sqrt{ {8 \pi \over 3} G\rho(t,r) x^2 + \frac{x^2[1+b(t)]}{a^2(t)}
}~.
\end{equation}
Further assuming that
the $t$-dependence of $x$ is not very different from that of
$a^2(t)/[1+b(t)]$ (since we integrate from 0 to 1 for $x$
and the $G\rho x^2$ term is proportional to $1/x$,
the calculation of the age is not sensitive to the last term
in the square root and the approximations used),
we obtain, from Eqs.(57) and (59),
\begin{equation}
t_0(r) \simeq \frac{1}{\overline{H}_0} \int_0^1 \frac{ \sqrt{x} dx }
{ \sqrt{ \overline{\Omega}_0 ( 1+ \delta(r) )+x(1-\overline{\Omega}_0)}}~~.
\end{equation}
Setting $\delta(r)=0$ in Eq.(62) reproduces the
age of the open Universe with $\overline{\Omega}_0 <1 $
in the Friedmann cosmology.
It is to be emphasized  that the true age of the Universe
is determined by $\overline{\Omega}_0$,
not by the global value of $\Omega_0$.
Since we expect $\delta(r)$ to be an increasing function of $r$
to be consistent with an increasing $\Omega_0(r)$,
$t_0(r)$ is a decreasing function of $r$.
The true age at $r\simeq 0$ ($\delta(r\simeq 0)=0$) is given by
\begin{equation}
\overline{t}_0 \simeq \frac{0.9}{ \overline{H}_0}=18 ~
\mbox {Gyr  for  } \overline{H}_0 = 50 \mbox{ Km/secMpc}~,
\end{equation}
where we have used $\overline{\Omega}_0=0.1$.
The above age can easily accommodate the observed ages, 14$\sim$18 Gyr
of the oldest stars and globular clusters in our galaxy.

\section{Summary and Conclusions}
We have proposed in this article a cosmological model
which accommodates isotropic but inhomogeneous matter distributions
as observed in the present Universe, at least up to scales of
several hundred Mpc.
In order to be consistent with this scale-dependence
in the metric which was motivated by the present inhomogeneous
matter distribution,
the Einstein equation was also generalized in such a way
that the energy-momentum tensor is also scale-dependent.
The scale-dependence of the gravitational constant, $G$,
was independently suggested by the effective
AFHD quantum gravity theory.

We have shown that the form of our metric severely constrains
the functional dependence of the scale factor, $R(t,r)$,
under a reasonable assumption that $p_r=p_{\theta}$.
In particular, it was shown that the new metric specifies
the $r$-dependence of $R(t,r)$.
In contrast to the standard Friedmann cosmology,
in which it is perfectly allowed to neglect the pressure
and the momentum density
compared with the matter density $\rho$
in the matter dominated era,
 it is essential, in our model, to keep $p$ and $T_{01}$,
even in the matter-dominated era,
as non-vanishing (no matter how small they are)
for $mathematical$ $consistency$.
(They can be negligible, at least in our local neighborhood, compared with
the matter density for physical applications.)
Otherwise, the $t$ and $r$ dependence is factorized
and the standard model is recovered.

In this model, due to the nature of the metric,
quantities such as $\rho$, $p$ and $T_{01}$
are completely determined by two functions of $t$,
$a(t)$ and $b(t)$.
Or, equivalently, knowledge of $\rho$, $p$ and $T_{01}$
determines $a(t)$ and $b(t)$.
The function $a(t)$ resembles the scale factor
in the standard cosmology and $b(t)$ controls the curvature
and the inhomogeneity of the matter distribution.
When  1 + $b(t)$ is positive, there emerges a horizon at
$r=2/\sqrt{1+b(t)}$ in this metric whose location changes in time.
In addition, in spite of the absence of the cosmological
constant in the Einstein equation,
a term which mimics the cosmological constant appears
and it is also $t$-dependent.
(In a special case, it becomes a constant.)
Because of this, the pressure can be negative in this metric.
We have demonstrated that by properly adjusting the
unknown but free $b(t)$ in this metric, we can reproduce,
at least qualitatively, the gross behavior of the Universe
from the early epoch to the present epoch.

The $r$-dependence introduced in this metric manifests itself
in the $r$-dependence of the expansion rate, $H_0$,
$\Omega_0$ and the age of the present Universe.
Of course, these quantities are not directly measurable.
Instead, physical quantities must be reduced from observables,
such as red shifts, $z$, luminosity distance, $d_L$,
and so on.
One of unfortunate features of our metric is that
the light propagation equation, which provides such relationships
among various observable  quantities,
cannot, in general, be integrated out to provide this vital
information in analytical forms.
However, the scale-dependence may significantly modify the standard relations
in the Friedmann cosmology, in particular, for large $z$.
(For small distance up to, say, of order of 100 Mpc,
the standard relationships hold with minor modifications of
the $r$-dependence.)

The local age of the present Universe, which is the true
age of the Universe in this model,  is given
in terms of the local density and expansion rate,
but with a caveat that the local expansion rate
is not well defined and thus unmeasurable.
This can only be inferred from the observed $H_0$ at various distances.
In our metric (with $\dot{b}(t) >0$), the expansion rate
and the density are increasing functions of the distance scale.
The scale-dependent feature is similar to the
running of the physical coupling constant such as
the fine structure constant, the strong coupling constant and
so on.
The values of these $constants$ depend on what energy scale one measures
them.

So far in this paper, we have studied
a generalization of the Robertson--Walker metric
by dropping the homogeneity.
It would be interesting to investigate other
possible modification of the Robertson--Walker metric.
The metric studied here is by no means unique.
Also, it is clear that further study on new dynamics
or new constraint is necessary to determine the
behavior of the $a(t)$ and $b(t)$.
Study of quantitative modifications of the standard Hubble law, for example,
due to the running of $H_{0}$, $\Omega_{0}$, and $t_{0}$, and
the derivation of $d_L$ as a function of $z$ and $r$ are  under way
and will be reported elsewhere.

\acknowledgments
The authors would like to thank G. Feldman who has
participated in many aspects of this  work  and has given us invaluable help
and advice.
They also wish to thank A. Chakrabarti, N. Fornengo,  G. Grumberg,
R.Holman, T. H. Lee, T.N.Pham and T.N. Truong for helpful
discussions.
This work was supported in part by the National Science Foundation.

\end{document}